%% file: main_sc_2025.tex
\documentclass[sigconf]{acmart}
\usepackage{enumitem}
\usepackage{algorithm}
\usepackage{algpseudocode}
\usepackage{amsmath}
\usepackage{amsthm}
\usepackage{mathtools}
\usepackage{subcaption}
\usepackage{graphicx}
\usepackage{cancel}
\usepackage{xcolor}
\usepackage{caption}
\usepackage{subcaption}
\usepackage{setspace}
\usepackage{placeins}
\usepackage{float}

\definecolor{bin}{RGB}{200, 000, 200}
\definecolor{dong}{RGB}{0, 0, 0}
\definecolor{sherry}{RGB}{255,140,0}

\definecolor{check}{RGB}{0,0,0}

\def\vn#1{\textcolor{magenta}{{#1}}} 

\usepackage{xspace}

\usepackage{setspace}

\copyrightyear{2025}
\acmYear{2025}
\setcopyright{none}
\acmConference[SC '25]{The International Conference for High Performance Computing, Networking, Storage and Analysis}{November 16--21, 2025}{St Louis, MO, USA}
\acmBooktitle{The International Conference for High Performance Computing, Networking, Storage and Analysis (SC '25), November 16--21, 2025, St Louis, MO, USA}
\acmDOI{10.1145/3712285.3759805}
\acmISBN{979-8-4007-1466-5/2025/11}

\AtBeginDocument{%
  }

\author{Bin Ma}
\orcid{0009-0003-4287-8946}
\affiliation{%
  \institution{University of California, Merced}
  \country{USA}
}
\email{bma100@ucmerced.edu}

\author{Viktor Nikitin}
\orcid{0000-0001-9999-169X}
\affiliation{%
  \institution{Argonne National Laboratory (ANL)}
  \country{USA}
}
\email{vnikitin@anl.gov}

\author{Xi Wang}
\orcid{0009-0001-6251-8177}
\affiliation{%
  \institution{University of California, Merced}
  \country{USA}
}
\email{swang166@ucmerced.edu}

\author{Tekin Bicer}
\orcid{0000-0002-8428-5159}
\affiliation{%
  \institution{Argonne National Laboratory (ANL)}
  \country{USA}
}
\email{tbicer@anl.gov}

\author{Dong Li}
\orcid{0000-0001-9336-0694}
\affiliation{%
  \institution{University of California, Merced}
  \country{USA}
}
\email{dli35@ucmerced.edu}

\renewcommand{\shortauthors}{Ma et al.}


\begin{document}
\newcommand{\name}{mLR\xspace}

\title{mLR: Scalable Laminography Reconstruction based on Memoization}

\renewcommand{\shortauthors}{Ma et al.}

\begin{abstract}
ADMM-FFT is an iterative method with high reconstruction accuracy for laminography but suffers from excessive computation time and large memory consumption. We introduce \textit{\name}, which employs memoization to replace the time-consuming Fast Fourier Transform (FFT) operations based on an unique observation that similar FFT operations appear in iterations of ADMM-FFT. We introduce a series of techniques to make the application of memoization to ADMM-FFT performance-beneficial and scalable. 
We also introduce variable offloading to save CPU memory and scale ADMM-FFT across GPUs within and across nodes. Using \name, we are able to scale ADMM-FFT on an input problem of $2K \times 2K \times 2K$, which is the largest input problem laminography reconstruction has ever worked on with the ADMM-FFT solution on limited memory; \name brings 52.8\% performance improvement on average (up to 65.4\%), compared to the original ADMM-FFT.
\end{abstract}


\maketitle
\input text/intro

\input text/background

\input text/algorithmic_fusion

\input text/memoization

\input text/scale_up_scale_out
\input text/evaluation
\input text/related_work
\input text/conclusion
\input text/acknowledgment

\bibliographystyle{ACM-Reference-Format}
\bibliography{li,shangye,bin,vn,sherry}


\end{document}

%% file: text/intro.tex
\section{Introduction} 

Laminography is a noninvasive, 3D imaging technique for studying flat or elongated structures with X-ray \cite{gondrom1999x,helfen2005high}. 
It has been widely deployed in scientific facilities, such as synchrotron light sources, to study the internal structures of materials and samples, including integrated circuits (IC), large-scale composite materials, and morphology of biological tissue~\cite{xu2010synchrotron,morgeneyer2014situ,reischig2013high}. 

During a laminography experiment, a sample is placed on a rotating stage and exposed to X-rays, while a detector captures a series of 2D projections from different angles. 
The laminographic reconstruction process (LR) is then used to reconstruct the 3D structure of the object from these projections. High-quality laminography reconstruction is essential for many scientific domains, as the reconstructed morphology and sample features serve as a foundation for accurate analysis and deeper insights into the scientific problem.
 However, unlike traditional tomography, 
laminography is prone to imaging artifacts caused by insufficient Fourier spectrum coverage~\cite{fisher2019laminography,nikitin2024laminography}, significantly limiting the quality of reconstructions. 
A recent study presented an accurate solution to this problem using Total Variation (TV) regularization and validated its performance on mouse brain specimens \cite{nikitin2024laminography}. 
The LR problem is solved iteratively using the Alternating Direction Method of Multipliers (ADMM)~\cite{boyd2011distributed}, 
with Fast Fourier Transform (marked as $F_{u*D}$)~\cite{beylkin1998applications,dutt1993fast} employed to accelerate the computation of forward and adjoint laminography. 



Although ADMM-FFT is an advance technique that can provide high-quality 3D reconstructions, its applicability to large-scale 3D volumes is limited due to the memory-intensive and computationally demanding operations. 
For example, the reconstruct of 1024$\times$1024$\times$1024 (or (1K)$^3$) volume by AMDD-FFT introduces more than 120GB CPU memory overhead, with the overhead growing cubically, following $O(N^3)$, where $N$ is the dimension size of the volume. This significantly hinders scientific progress that relies on 3D X-ray imaging, particularly at synchrotron radiation facilities, where extremely large 3D volumes are encountered---such as reconstructing a brain sample at sub-micron resolution or an IC at sub-10-nm resolution \cite{hidayetouglu2019memxct,hidayetoglu2020petascale}. 

In this paper, we introduce \name to accelerate ADMM-FFT and scale the workload across multiple GPUs to address the above problems. To reduce the expensive computation time, \name uses a novel memoization technique. This technique is based on our unique observation that across iterations of ADMM-FFT, the inputs to the FFT operations can show similarity, leading to similar results in the FFT operations. The existence of such similarity allows us to apply the memoization technique, in which we store and reuse the results of the FFT operations during iterations. Applying the memoization, we can avoid both expensive computation and data transfer between the CPU and GPU, leading to an order of magnitude reduction in the execution time of the FFT operations, compared with the original GPU-based implementation. 

However, applying the memoization to the FFT operations, we face two challenges. \textit{First}, we face a fundamental tradeoff between providing large memory capacity to the memoization and diminishing performance return from the memoization. In particular, applying the memoization, we store each pair of operation input and output as a key-value pair in a memorization database, which can easily consume TBs-scale storage space for certain input problems. To accommodate such a large database, we can use a memory node that remotely provides large memory capacity. However, frequently interacting with the memory node for data retrieval diminishes the performance benefit brought by the memoization. \textit{Second}, searching the memoization database using the high-dimensional keys (i.e., the operation inputs) to find matching value can be expensive. The dimensionality of the operation input is often at \textcolor{check}{the dimensions often at the order of $10^3$ elements.} Calculating the distance between two keys is expensive. 

To address the first challenge, \name introduces operation cancellation and fusion, and memoization caching. The operation cancellation and fusion numerically transform the original sequence of FFT operations into an equivalent shorter sequence without impacting the program correctness. Such transformation reduces the accesses to the memory node by 33\%. The memoization caching duplicates some key-value pairs in the local memory of the compute node (running ADMM-FFT), such as the accesses to the memory node are reduced. The memoization cache is a small private cache to save memory overhead while saving the cache management overhead. 
In particular, the cache is private to each chunk (i.e., a location in the input image), and used to cache recent value retrieved from the memoization database for that chunk; the cache is not shared across all chunks as a global cache. Our cache design is driven by our observation that using a private cache can achieve the similar cache hit rate as a global cache but saving the computation needed for caching by 85\% (compared to using the global cache).  

To address the second challenge on the high-dimensionality of the keys, we introduce a CNN-based encoder to map the input (the key) to a lower-dimensional space. The encoder allows us to efficiently find similar keys while capturing the nature of the image. 

To enable ADMM-FFT on much larger input problems, \name introduces ADMM-Offload. This technique offloads memory-consuming variables to SSD. ADMM-Offload prefetches variables to the CPU memory before their accesses in the next execution phase such that the data movement overhead is hidden.  To determine the prefetch distance (i.e., when to trigger prefetch), 
we analyze the repetitiveness of memory access patterns and form offloading/prefech plans to maximize memory saving and minimize data movement overhead. Using ADMM-FFT plus the chunk distribution across nodes in \name, we scale ADMM-FFT on a dataset of $2K \times 2K \times 2K$, which is \textit{the largest LR problem that the domain scientists have ever worked on with the ADMM-FFT technique} on memory-constrained nodes.

Our main contributions are summarized as follows. 

\begin{itemize}
    \item We characterize the FFT operations (the major computation) in ADMM-FFT and reveal the unique memoization opportunities to accelerate it;
    \item We customize the memoization for ADMM-FFT based on FFT operation transformation and caching to maximize the performance return of the memoization; We improve the scalability of ADMM-FFT within a node and across nodes;
    \item Our evaluation reveals that \name brings 52.8\% performance improvement on average (up to 65.4\%) and enables ADMM-FFT on large input problems, providing new opportunities for scientific discovery. \textcolor{dong}{Our code is open-sourced}~\footnote{https://github.com/anonimo-v/OpenLB.git}.
\end{itemize}

%% file: text/background.tex
\section{Background} 
\label{sec:background}


Laminography reconstruction (LR) with regularization can be regarded as solving an optimization problem that balances data fidelity and smoothness, formulated in Equation~\ref{eq:target objective}. 
\vspace{-5pt}
\begin{equation}
\label{eq:target objective}
\min_{u} \frac{1}{2}\| L u - d \|_2^2 + \alpha \|  u \|_\text{TV}
\end{equation}

The above optimization problem contains a data fidelity term ($\min_{u} \frac{1}{2}\| L u - d \|_2^2$) and a regularization term ($\alpha \|  u \|_\text{TV}$). The data fidelity term maps a reconstructed object $u$ to observed data $d$ by applying FFT operations $L$ to $u$, and minimizes the difference between $d$ and $Lu$. The regularization term, particularly the Total Variation (TV), helps reduce noise and artifacts.
This optimization problem can be accurately solved using the Alternating Direction Method of Multipliers (ADMM), in which the input problem is split into simpler subproblems.





\textbf{ADMM} is effective to solve problems with constraints~\cite{boyd2011distributed}. ADMM is used to iteratively update variables and multipliers to satisfy some constraints while optimizing an objective function. In the context of LR with TV,  Equation~\ref{eq:target objective} is augmented into a Lagrangian form, and \textcolor{dong}{ADMM is applied to this form to solve LR with TV.}



ADMM splits the optimization problem into two subproblems, 
the laminography subproblem (LSP) and regularization subproblem (RSP), while  incorporating iterative parameter updates. 
LSP updates (or refines) the main reconstruction image $u$ using a small number of conjugated gradient (CG) iterations ($N_{inner}$);  
RSP updates an auxiliary variable $\psi$, which is related to the TV regularization term $\alpha \|  u \|_\text{TV}$. There are additional parameters $(\rho, r,$ and $ s)$ updated to accelerate ADMM convergence. 

RSP is computationally lightweight. 
In contrast, LSP is more computationally demanding. We discuss LSP in detail in the remaining section. The variable $\lambda$ is used to coordinate the two subproblems to ensure consistency when ADMM is applied. 
\begin{algorithm}[t]
\caption{LSP}
\label{alg:lsp_fft_left}
\begin{algorithmic}[1]
\Require data $d$, initial guess for $u$, fixed variables $(\psi, \lambda, \rho)$, inner ADMM iterations $N_\text{inner}$.
\State $g \gets \psi-\lambda/\rho$, $G_\text{prev} \gets 0$
\For{$i = 1$ to $N_\text{inner}$}
\State {\bfseries Forward Pass:}
\State{$d' \gets F^*_{\text{2D}}F_{\text{u2D}}F_{\text{u1D}} u$}
\State{$g' \gets \nabla u$}
\State {\bfseries Adjoint Pass:}
\State{$G \gets F^*_{\text{u1D}}(F^*_{\text{u2D}}(F_{\text{2D}} (d'-d)))-\rho\nabla
(g'-g)$}
\State {\bfseries CG update:}
\State $u\gets CG(u,G,G_\text{prev})$
\State $G_\text{prev}\gets G$
\EndFor
\State \Return $u$
\end{algorithmic}
\end{algorithm}

\begin{figure}[t!]
    \centering
        \includegraphics[width=\columnwidth]{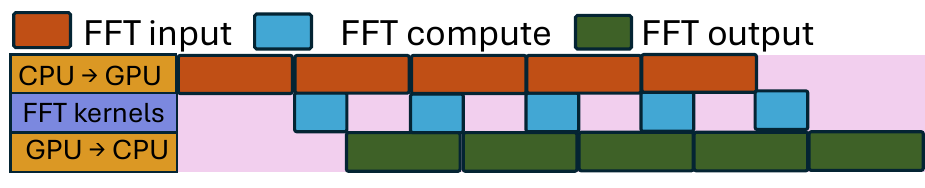}
    \vspace{-15pt}
    \caption{Computation and communication pipeline for the operation $F_{u1D}$ in the existing work.}  
    \label{fig:usfft1d_timeline}
    \vspace{-10pt}
\end{figure}


\textbf{Fast Fourier Transform} (FFT). ADMM works with FFT-based methods to efficiently solve LSP. Those methods have shown great promise in addressing LR challenges due to their computational efficiency for large datasets, and have potential to alleviate issues related to data size~\cite{nikitin2024laminography}. Moreover, they yield more accurate and stable solutions when combined with regularization (e.g., TV).

Algorithm \ref{alg:lsp_fft_left} summarizes the key steps for solving LSP, employing FFT with the operations $F_\text{u1D}$, $F_\text{u2D}$ and $F_\text{2D}$ and their inverse $F^*_\text{u1D}$, $F^*_\text{u2D}$ and $F^*_\text{2D}$. 
The operation $F_{\text{2D}}$ represents the two-dimensional Fourier transform on equally spaced grids, whereas the operations $F_{\text{u2D}}$ and $F_{\text{u1D}}$ denote the two- and one-dimensional Fourier transforms on unequally spaced grids. 

\begin{figure}[t!]
    \centering
        \includegraphics[width=\columnwidth]{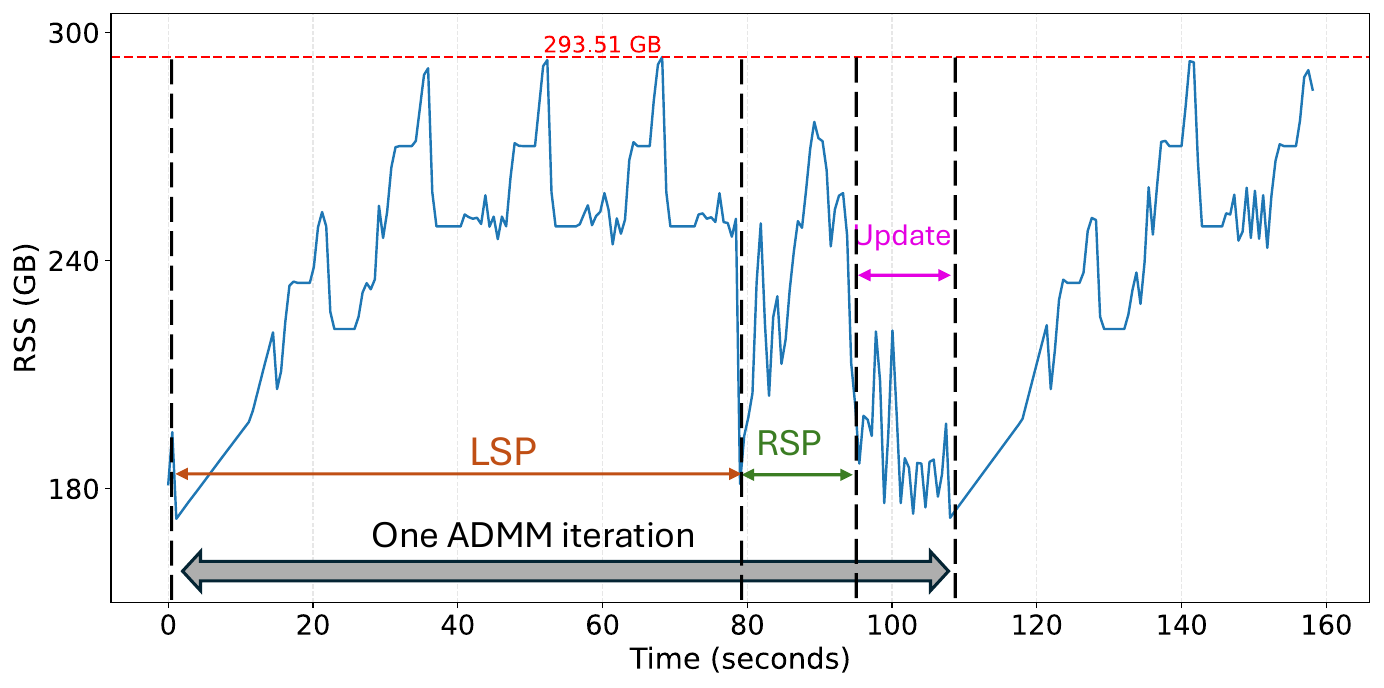}
    \vspace{-15pt}
    \caption{CPU memory consumption in one ADMM iteration.}
    \label{fig:mem_use}
    \vspace{-15pt}
\end{figure}


\textbf{CPU memory consumption in ADMM.} ADMM has a large memory footprint due to its auxiliary variables and intermediate computation results. For example, reconstructing a 3D volume
from 1.5K projections (each with 1.5K $\times$ 1.5K dimensions) requires $\sim$300 GB of memory, which is 
25 $\times$ more than the input size of 11.4 GB. 
Figure~\ref{fig:mem_use} shows CPU memory consumption in one ADMM iteration for this input problem. 
The auxiliary $\psi$ and Lagrange multipliers $\lambda$ use 34 GB (12\%) of memory each, and the gradient ($g$) plus the gradient from the previous iteration ($g_{prev}$) require 68.4 GB (24\%).  


\textbf{LSP} 
dominates the overall execution time of ADMM. Figure~\ref{fig:mem_use} gives an example where LSP accounts for more than 67\% of the total time. LSP is dominated by frequent invocation of FFT operations. 


In LSP, the large input dataset is broken into smaller \textit{chunks} to fit into the GPU memory \cite{nikitin2024laminography}. A chunk is a partition of an input 3D array (often transformed into the frequency domain) along a specific dimension, and the operations $F_{u1D}, F_{u2D}, F^{*}_{2D}, F_{2D}, F^{*}_{u2D}, and F^{*}_{u1D}$ work on a single chunk at any given time.  Specifically, the input dataset is initially loaded into host memory (CPU memory), and then the chunks are fetched to GPU memory one by one. To hide the chunk transfer overhead, the existing approach 
overlaps chunk movement and FFT operations.
\textcolor{check}{
For clarity, the input/output dimensionalities of the key operators are as follows:
\[
u \in \mathbb{R}^{(n_1, n_0, n_2)}, \quad d \in \mathbb{R}^{(n_{\theta}, h, w)},
\]
\[
F_{u1D}: u[n_1, n_0, n_2] \;\longrightarrow\; \tilde{u}1[n_1, h, n_2],
\]
\[
F_{u2D}: \tilde{u}1[n_1, h, n_2] \;\longrightarrow\; \tilde{u}2[n_{\theta}, h, w].
\]
The corresponding adjoint operators have inverse input/output dimensionality.}

Figure~\ref{fig:usfft1d_timeline} depicts the execution pipeline for one FFT operation. 


Despite the pipeline optimization, the data transfer between GPU and CPU still causes significant overhead. For example, when the input problem is $1K \times 1K \times 1K$, the data transfer exposed to the critical path accounts for 47\% of the total execution time.

%% file: text/algorithmic_fusion.tex
\section{Overview}
\name uses the memoization technique to replace expensive FFT operations in LSP. To accommodate a  large memoization database, \name uses a distributed  memoization system where a pair of compute node (with the GPUs) and memory node are used to apply the memoization. To maintain the quality of the final output of ADMM-FFT after applying memoization, we use a threshold $\tau$ to control the frequency of applying memoization based on domain knowledge. \textcolor{dong}{Note that \name does not change the FFT algorithm; instead it reduces the number of calls to FFT operations with memoization to improve overall performance. Hence, \name can work with and complement FFT algorithms.} 

To reduce frequent accesses to the remote memory node for memoization, \name introduces operation cancellation to reduce FFT operations. However, this brings the LSP computation to the frequency domain, increasing computation complexity on the CPU (especially a subtraction operation on the CPU). To avoid such a problem, we run the loss calculation in LSP on the GPU and fuse it with neighboring FFT operations. 

Furthermore, the memoization in \name uses a CNN model as an input encoder to reduce the input dimensionality and speed up the similarity search. Building such a CNN model is challenging because of the lack of labeling to quantify the similarity between training samples, we employ the contrastive learning to solve this problem. In addition, \name adds a memoization cache on the CPU node to reduce frequent accesses to the memory node. Such a cache is private to each chunk location in the input image in order to save the computation overhead of caching while keeping the output quality of ADMM-FFT after using memoization. 

To scale ADMM-FFT on larger input problems on limited CPU memory, \name uses ADMM-Offload that offloads variables from the CPU memory to SSD to save the CPU memory. The selection of the variables for offloading and prefetch is guided by four constraints that determine when to offload and when to prefetch variables at execution phases. In addition, we introduce a metric during the variable selection process to strike a balance between saving memory and reducing performance loss caused by data movement (i.e., offloading and prefetch).  To scale ADMM-FFT across GPUs, \name uses the chunk-based input partition method in ADMM-FFT to distribute workloads. \textcolor{dong}{Note that by default the term ``performance'' means execution time in this paper.}

%% file: text/memoization.tex
\section{Memoization}
\label{sec:memoization}
\begin{figure}[t]
\centering
\captionsetup[]{justification=centering}
    \includegraphics[width=1\columnwidth]{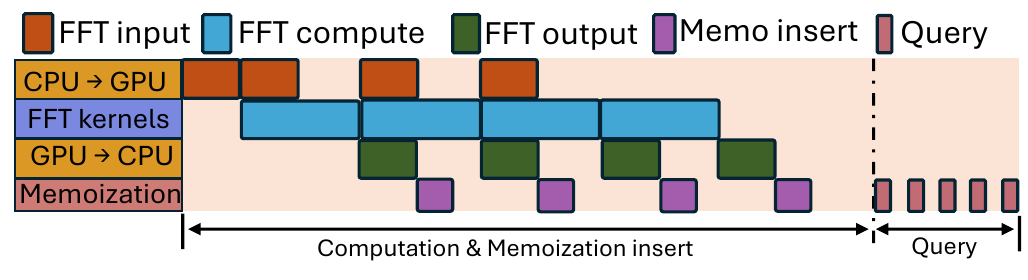}
\caption{mLR's execution pipeline for the operation $F_{u2D}$ with memoization}
\label{fig:memoization_pipe}
\vspace{-13pt}
\end{figure}

\textcolor{check}{We present a distributed memoization system to accelerate ADMM-FFT by replacing expensive FFT operations (which consume >50\% of iteration time) with cached results.} 
Figure~\ref{fig:memoization_pipe} shows the general memoization timeline. 

\subsection{Memoization Object}

The use of memoization must meet two criteria. First, the computation to be replaced, referred to as the memoization object, must be sufficiently longer than the memoization overhead ---including data retrieval and searching in the memoization database--- to ensure a performance gain. Second, the approximation error introduced by the memoization must be tolerable by the application with acceptable impacts on computation correctness.

\begin{figure}[t!]
    \centering      
    \includegraphics[width=1\columnwidth]{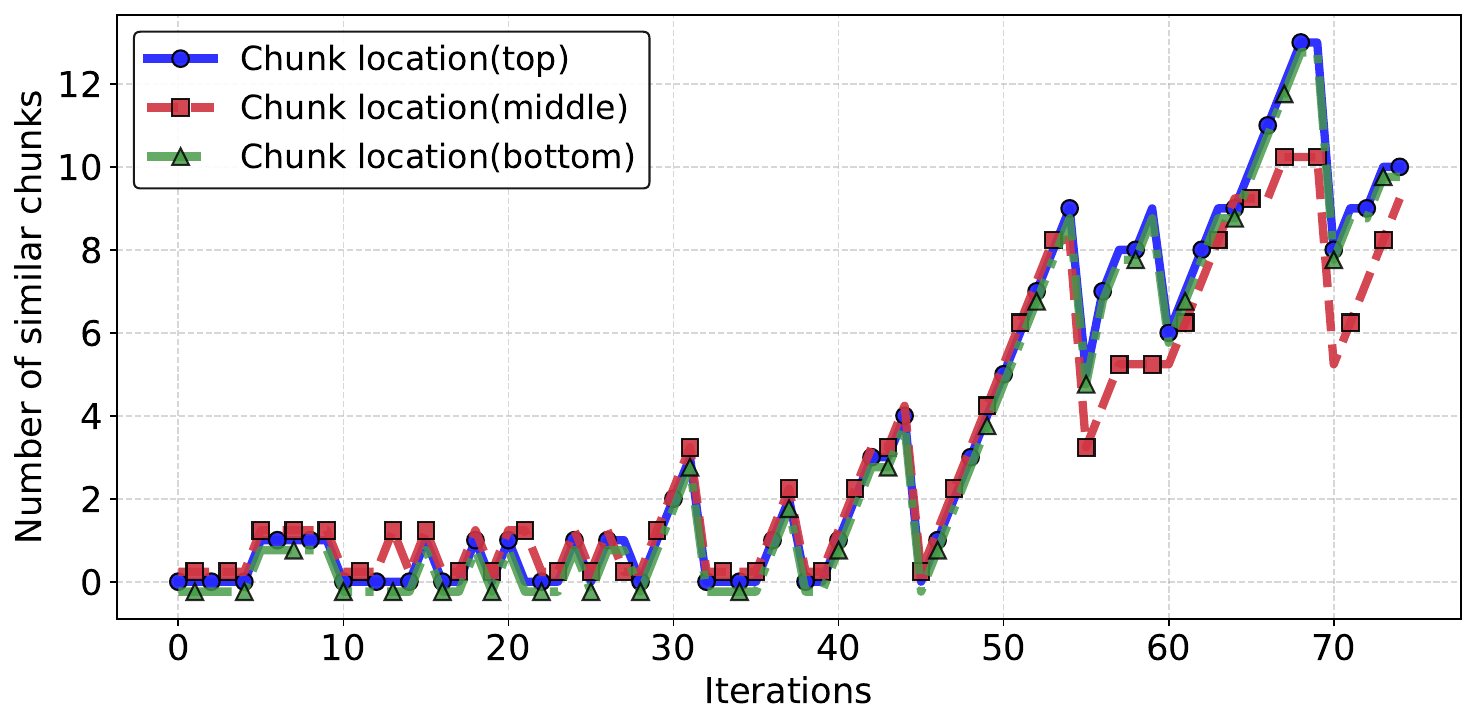}  
        \vspace{-15pt}
        \caption{At a chunk location, similar chunks can appear across the iterations of ADMM-FFT.}
        \label{fig:simi_chunks}
        \vspace{-12pt}
\end{figure}

\begin{algorithm}[t]
\caption{Optimized LSP. This is an extension to Algorithm 2. \textit{The extension is highlighted in shaded statements.}}
\label{alg:lsp_fft_right}
\begin{algorithmic}[1]
\Require data $d$, initial guess for $u$, fixed variables $(\psi, \lambda, \rho)$, inner ADMM iterations $N_\text{inner}$.
\State $g \gets \psi-\lambda/\rho$, $G_\text{prev} \gets 0$
\State  \colorbox{lightgray}{$\hat{d} \gets F_{\text{2D}}d$}
\For{$i = 1$ to $N_\text{inner}$}
    \State {\bfseries Forward Pass:}
    \State{$\hat{d'} \gets \colorbox{lightgray}{$F_{\text{u2D}}(F_{\text{u1D}} u)$}$}
    \State{$g' \gets \nabla u$}
    \State {\bfseries Adjoint Pass:}
    \State{$G \gets \colorbox{lightgray}{$F^*_{\text{u1D}}(F^*_{\text{u2D}} (\hat{d'} -\hat{d}))$} - \rho\nabla^(g'-g)$}
    \State {\bfseries CG update:}
    \State $u\gets \text{CG}(u,G,G\text{prev})$
    \State $G\text{prev}\gets G$
\EndFor
\State \Return $u$
\end{algorithmic}
\end{algorithm}

We choose to replace the basic computing element, FFT operations in Algorithm \ref{alg:lsp_fft_left}, in particular, three FFT operations in Line 4 and three FFT operations in Line 7, with memorization. Three key observations motivate this optimization, as outlined below. 

\begin{itemize}[leftmargin=*]
    \item The 1D and 2D FFT operations in Algorithm \ref{alg:lsp_fft_left} are expensive and take at least 50\% of the iteration time.  Those operations happen on GPU, and come with data transfer between CPU and GPU before the operations. Replacing those operations with memoization can avoid \textit{both} expensive computation and data transfer, which is performance beneficial. 
     
    \item Iterative nature of Algorithm \ref{alg:lsp_fft_left} allows us to tolerate computation approximation in the gradients introduced by memoization. 

    \item Each forward pass of FFT 1D or 2D operation takes a chunks as input and then outputs a chunk of estimated data object $\hat{d'}$ (a 3D array). 
    Explicitly defined input and output, which can be used as the key and value to the memoization database respectively, enables straight application of memoization. Also, according to our profiling results, reading/storing input and output data objects from/to CPU memory due to  memoization is one order of magnitude cheaper than transferring them to GPU and compute.  
\end{itemize}


Furthermore, we observe that at each chunk location, the input chunk across iterations can show similarity defined in terms of the cosine similarity (discussed more in Section~\ref{sec:correctness}). The existence of such similarity allows us to store results of the FFT operations for a chunk location to be reused in future iterations. \textcolor{dong}{Such similarity arises because of the iterative nature of ADMM-FFT. As ADMM-FFT is converging, the update to a chunk location across the iterations becomes smaller and smaller.}

To demonstrate the above observation, we show the chunk similarity for three chunk locations across iterations in Figure~\ref{fig:simi_chunks}. In the figure, ADMM-FFT uses a downsampled mouse brain dataset with a dimension of $1K^3$.  The whole workload runs 75 iterations. There are 125 chunk locations in this example, and the figure shows the results for three representatives (i.e., the 1st as the top location, the 62nd as a middle location, and the 125th as the bottom location). During the evaluation, we introduce a parameter $\tau$ to determine the similarity between two chunks. If the cosine similarity between two chunks is larger than $\tau$, then they are similar. $\tau = 0.93$ in our evaluation, but we have the same observation when $\tau \in [0.9, 0.95]$. We have more discussions on $\tau$ in Section~\ref{sec:correctness}. 

Figure~\ref{fig:simi_chunks} shows that chunk similarity across iterations commonly exists. For example, after 30 iterations, a chunk location can find 4-9 similar chunks in prior iterations. In 70\% of all iterations, we are able to find similar chunks in prior iterations. Also, as we finish more iterations, we are able to find more similar chunks in prior iterations. For example, in the iteration 59, we find 9 similar chunks.

In general, we replace the six FFT operations with the memoization,    \textcolor{dong}{which does not rely on any specific optimizations within FFT. We do not change FFT implementation.} To minimize the memoization overhead, we introduce FFT operation cancellation and fusion, such that we can reduce the frequency of applying memoization. 


\subsection{Operation Cancellation and Fusion} 
\begin{figure}[t]
\centering
\vspace{-10pt}
\captionsetup[]{justification=centering}
\subfloat[LSP after operation cancellation (no operation fusion)]{%
    \includegraphics[width=1\columnwidth]{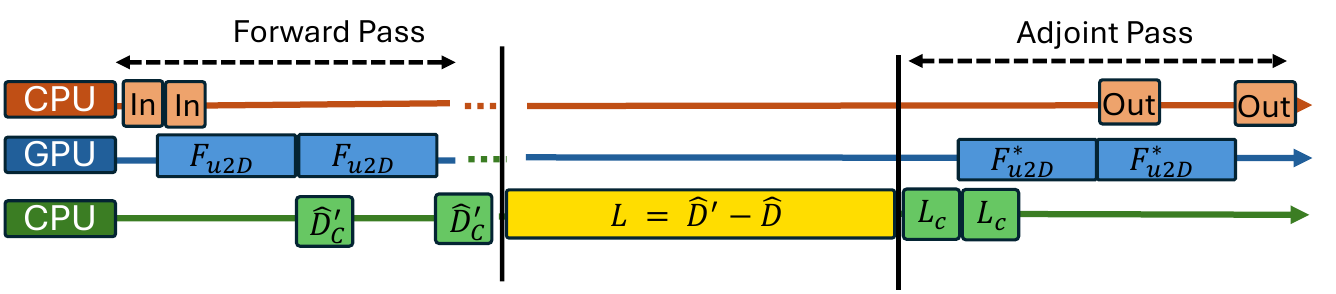}
    \label{fig:fusion_sub1}
}\\ %
\subfloat[LSP after operation cancellation and has operation fusion]{%
    \includegraphics[width=1\columnwidth]{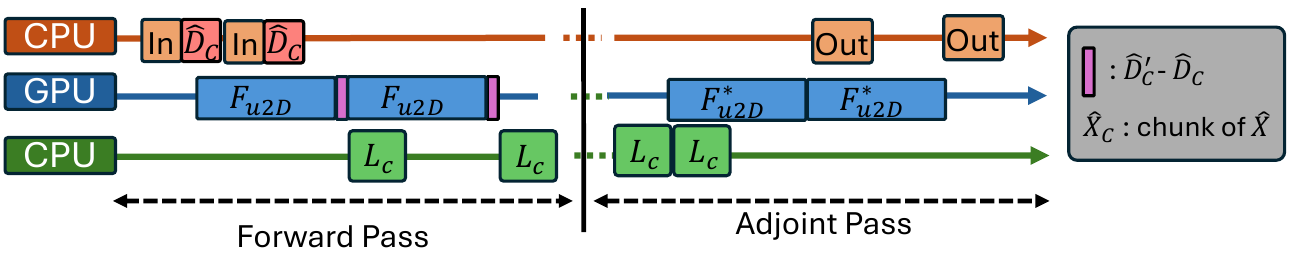}
    \label{fig:fusion_sub2}
}
\caption{LSP with and without operator fusion. }
\label{fig:operator_fusion}
\vspace{-15pt}
\end{figure}

We cancel and fuse the FFT operations without affecting the correctness of the execution (see Figure~\ref{fig:operator_fusion}). The operation cancellation changes the original three-step FFT reconstruction to a two-step method (Lines 2 and 3 in Algorithm \ref{alg:lsp_fft_right}). This method is enabled by mapping the input to the frequency domain. In particular, we perform $F_{2D}$ on the spatial domain data once it is loaded.


Based on the input mapping, we cancel operations as follows. In Algorithm \ref{alg:lsp_fft_left}, we discover that the operator $F^*_{2D}$ (Line 4 in Algorithm~\ref{alg:lsp_fft_left}) and its inverse $F_{2D}$ (Line 7 in Algorithm~\ref{alg:lsp_fft_left}) can be canceled because of their identity relationship (i.e., $F_{2D}F^*_{2D} = I$). In addition, $F^*_{2D}$ in the forward stage and $F_{2D}$ in the adjoint stage preserve the input/output dimensionality after the forward and adjoint, which ensures the correctness of the computation after operation cancellation. Note that we cannot cancel $F_{u1D}$, $F^*_{u1D}$, $F_{u2D}$, and $F^*_{u2D}$ based on the identity relationship because their input/output dimensionality is not preserved, leading to incorrect computation after operation cancellation. Operation cancellation eliminates $F_{2D}$ and $F^*_{2D}$, thereby reducing CPU-GPU data transfer time by 1/3.

After operation cancellation, the element-wise subtraction operation ($\hat{d}'- \hat{d}$) (shown at Line 8 in Algorithm \ref{alg:lsp_fft_right}) happens in the frequency domain instead of in the space domain as Algorithm \ref{alg:lsp_fft_left}, which increases the computation complexity, as the frequency domain uses \texttt{COMPLEX64} and the space domain uses \texttt{INT64}. As a result, running the subtraction operation on the CPU negates the gains from operator cancellation, and  \textcolor{dong}{extends execution time} by 5.1\%  in our evaluation on the dataset $1K \times 1K \times 1K$. 


To address the above problem, we run the subtraction on GPU (see  Figure~\ref{fig:fusion_sub2}). Also, we fuse this operation with $F_{u2D}$ on GPU to reduce kernel launch overhead. 
\textcolor{dong}{The fusion is implemented by adding the subtraction's input as a new argument to the FFT. Within the fused kernel, this new input is subtracted from the FFT output as the output of the fused code. We add 30 lines of code for fusion.}

However, running the subtraction operation on GPU leads to a transfer of the mapped data ($\hat{D_c}$ in Figure \ref{fig:fusion_sub2}) from  CPU to GPU. The overhead of data movement can be hidden. In particular, $F_{u2D}$ for a chunk is much longer than the data movement time for another chunk, hence overlapping with data movement.







\subsection{Distributed Memoization}
Our memoization mechanism in essence is a key-value store. It takes the input to the FFT operation as the key to search a memoization database. As a result, a data object similar to the FFT operation result is returned from the memoization database as the value.  

\subsubsection{Key Encoding} 
We do not directly use the input of the FFT operation to query the memoization database. Instead, we encode the input of the FFT operation using a convolutional neural network (CNN), which reduces the input dimensionality and accelerates the similarity search. We place the CNN model on the CPU to avoid the CPU-GPU data movement over PCIe. 
\textcolor{dong}{We apply INT8 quantization to the weights of the CNN model, and optimize its performance using vectorization (AVX512 instructions). The CNN inference time on the CPU takes less than 1\% of the total execution time according to our evaluation, which is very small.}


\textbf{Why CNN as the encoder?} We use CNN because of its superior performance in processing structured data, such as the input to $F_{u*D}$,  which is a \textit{chunk} of layered image in the frequency domain. Compared to the traditional image hashing \cite{image_hash2006} and multilayer perceptron (MLP), CNN encoding can extract more distinctive features because it  fuses information from multiple receptive fields across network layers. As a result, the chunks with similar frequency patterns can be encoded closer in the low-dimensional space, which is 
essential for query performance with memoization. 
Compared to the transformer architecture \cite{AttAllNeed17} , the CNN encoder can process high-dimensional arrays significantly faster.  


\textbf{CNN architecture.} Our CNN has three layers. The first layer has 32 filters, each with the size of $5\times5$. The second layer has 64 filters, each with the size of $3\times3$. The third layer is a fully connected layer which embeds the features extracted by the prior layers into a lower-dimensional space. 


\textbf{CNN input.} The input to the FFT operation is a \texttt{COMPLEX64}-typed matrix,  in contrast, the traditional CNN implementations typically support computation with (potentially lower precision) floating-point numbers. Consequently, the AI frameworks, such as PyTorch and TensorFlow, do not support the construction of \texttt{COMPLEX64}-typed CNN. To address this problem, the \texttt{COMPLEX64}-typed matrix is decomposed into two matrices, corresponding to the real and imaginary components of the matrix element, respectively. Mathematically, this method preserves both the magnitude and phase information inherent in the \texttt{COMPLEX64} representation. In addition, the matrix decomposition is lightweight and captures the complete frequency information embedded in the input.

\textbf{CNN training.}
\textcolor{dong}{The CNN is trained on GPU.} The goal of the training is to encode the input chunks such that ``similar'' chunks have a ``similar'' vector-based representation in a low-dimensional space. To reach the above goal, given a training sample (a chunk), we need to find another chunk to calculate their similarity. We employ the contrastive learning method \cite{contrastive_L2020}. Using this training method, we input two chunks in each training iteration, and each chunk is fed to the CNN encoder. Then, we calculate the training loss using the following equation. 

\begin{equation}
\small
\mathcal{L} = \left| \: \lVert\mathbf{z}_a - \mathbf{z}_b\rVert_2 - \lVert\mathbf{Ch}_a - \mathbf{Ch}_b\rVert_2 \: \right|
\end{equation}


The term $ \lVert\mathbf{Ch}_a - \mathbf{Ch}_b\rVert_2 $ defines the L2 norm of two input chunks $\mathbf{Ch_a}$ and $\mathbf{Ch_b}$, which is used as the ground-truth label. The term $ \lVert\mathbf{z}_a - \mathbf{z}_b\rVert_2 $ defines the L2 norm of two CNN outputs. 




\subsubsection{Distributed Memoization} 
\label{sec:dm}
We introduce the memoization mechanism. In general, this mechanism efficiently processes and stores the results of FFT operations from prior iterations and then tries to reuse them for future iterations to replace the computation. 

The results of FFT operations from previous iterations are stored in a memoization database. When local CPU memory is insufficient, this database is built on a remote memory node. 
The compute and remote memory nodes are connected by high-performance interconnect providing high bandwidth (hundreds of GB/s). This bandwidth is larger than that of local SSD (a few GB/s). We call such an architecture, distributed memoization.  Distributed memoization replaces the slow NVMe-based CPU-GPU data transfers (shown in Figure \ref{fig:memoization_pipe}) with faster (coelesced) inter-node communications, improving the performance.

\begin{figure}[t!]
    \centering      
    \includegraphics[width=1\columnwidth]{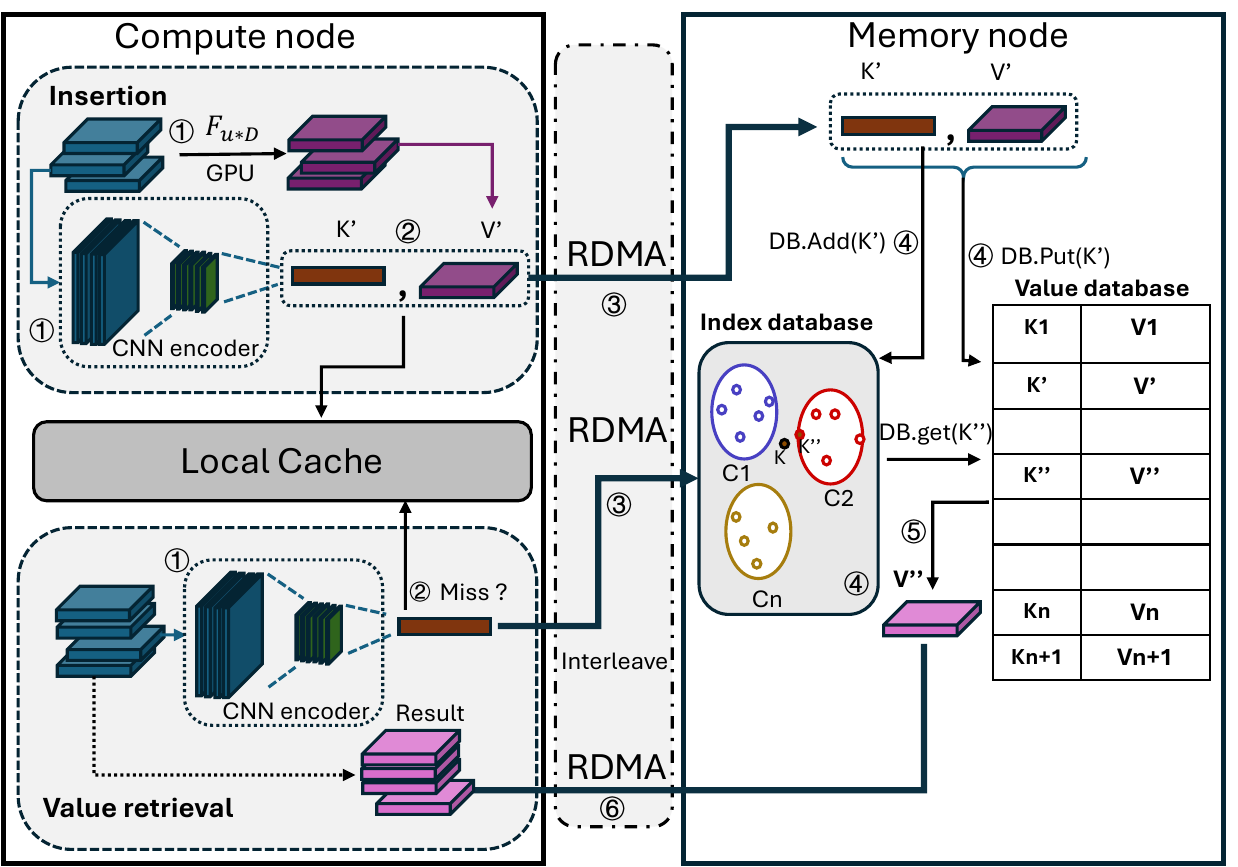}  
        \caption{\textcolor{check}{The scheme of distributed memoization.}}
        \label{fig:memoization}
        \vspace{-10pt}
\end{figure}
\textcolor{check}{Figure~\ref{fig:memoization} depicts the workflow of distributed memoization.} For each FFT operation, the compute node encodes the operation input (a chunk) as the key to query a remote memory node that stores precomputed results. If there is no match, the  compute node performs the FFT operation on the GPU and the computation result, along with the encoded key, is stored in the memory node. The above workflow is called \textit{insertion}.  In addition to the FFT operation time, the insertion introduces the following overheads: (1) key encoding, (2) database query, and (3) store of FFT operation result. The overhead (3) is hidden by using asynchronous store, while the overhead (1) and (2) are in the critical path. 

If a match value is found in the memoization database, it is returned to the compute node as the result of the FFT operation. This process, called \textit{value retrieval}, replaces the compute-intensive FFT operation with cheaper key encoding and database query.

The memory node hosts the memoization database, including an index (key) database and a value database. The index database organizes the keys based on their similarity. Querying the index database, we retrieve a key from the index database which is the most similar to the input query. The similarity is determined in terms of the L2 norm. The key retrieved from the index database is controlled by a threshold $\tau$. Only when the distance between the query key and the stored key is smaller than $\tau$, the stored key is returned. Hence, it is possible that no similar key can be found and retrieved from the index database. Using the returned key from the index database as an index, we find the corresponding value (the result of FFT operation) from the value database.

The index database is built using Faiss \cite{douze2024faisslibrary}, which uses an approximate nearest-neighbor (ANN) search algorithm to efficiently locate the most similar key. The Faiss' ANN provides two options for key organization: cluster-based (e.g.,  Inverted File indexing \cite{IVF_index}) and graph-based (e.g., Hierarchical Navigable Small Worlds indexing \cite{HSNW}). We use the cluster-based ANN in Faiss because it allows dynamic insertion with minimal overhead compared to the graph-based ANN, which incurs high reconstruction costs. Querying the index database is fast: Given an index database with one million keys with a key dimensionality of 60, querying the index database on our platform (AMD EPYC 7713 64-Core CPU and 512GB DDR4 memory) only takes 0.2 $ms$, $100\times$ shorter than the longest FFT operation for a chunk (i.e., $F_{u2D}$). 

The value database is built using Redis\cite{Redis2025}, a high performance in-memory database offering low latency and high throughput value retrieval. On our platform, the P99 latency of querying the value database is below 0.5 $ms$.

\subsubsection {Optimization of Payload Size}
\label{sec:op_payload_size}

For each memoization query, a key transfer  (less than 1 KB) needs to be performed between compute and memory nodes, achieving low utilization of interconnect bandwidth. To improve utilization, we introduce key coalesce. In particular, on the compute node, we add a small buffer to accumulate keys that require database queries. The  keys are accumulated until the communication payload size reaches 4 KB. This size leads to 95\% of bandwidth utilization in our platform using HPE Slingshot 11 with bidirectional injection bandwidth of 200 Gb/s. 

The key coalesce occurs across chunks but not within a chunk. Within a chuck, there are four FFT operations (after operation cancellation and fusion) with dependency between each other as shown in Algorithm~\ref{alg:lsp_fft_right}. Coalescing their keys invalidates program correctness. Across chunks, the keys are independent, and buffering them for batch processing improves bandwidth utilization.

Besides improving bandwidth utilization, key coalesce benefits performance from the following two perspectives. First, key coalesce enables batched lookup in the index database. Using multi-threading, batched lookup makes the best use of memory bandwidth in the memory node. Second, key coalesce amortizes performance overhead such as RDMA connection setup. 

\subsection{Memoization Cache}
To avoid frequent accesses to the memory node, the compute node maintains a cache, called the memoization cache, on the CPU memory of the compute node, which stores some values returned from the memoization database. Each item in the cache is a vector (a value) plus the corresponding key. Once a key is generated from ADMM-FFT, the compute node checks the cache using that key (resulting in a cache hit/miss). Only when the distance between that key and a key associated with a cache item is smaller than the threshold $\tau$, the cached value is used.

The memoization cache is a private cache, which means that each input chunk location has a cache. We do not build a global cache shared across all chunk locations for the following two reasons. First, the global cache and the private cache lead to the similar cache hit rate (shown in Section~\ref{sec:fft_cache}); Second, the overhead of the private cache is much smaller than that of the global cache. To determine the cache hit in the private cache, the similarity comparison we need to do is much smaller than that of using the global cache, because of the smaller capacity of the private cache. In our evaluation, the private cache saves the computation (similarity comparison) by 85\%, compared to using the global cache.




The memoization cache is small. For each chunk location, its private cache has only one item. We use this small cache size, such that the overall cache size for all chunks is equal to the original output size in ADMM-FFT, making memory consumption manageable. 


The memoization cache uses First-in-First-Out (FIFO) as the cache replacement policy. 
This means that once a value is fetched from the memoization database, it will replace the item in the corresponding cache. This is based on our observation that the cache has great temporal locality: the same chunk location across neighbor iterations tend to have the similar results in an FFT operation.

\subsection{Correctness and ADMM Convergence}
\label{sec:correctness}
\textbf{Correctness.} The quality of the reconstructed 3D object/image, i.e., the final output of ADMM-FFT, is sensitive to memoization, since memoization approximates the FFT computation. ADMM-FFT can tolerate a certain level of computation error due to its iterative nature \cite{Aslan2019,lossIterTao18}. 
However, to maintain the desired reconstruction quality, the value retrieved from the memoization database must be sufficiently close to the actual FFT result, ensuring that the approximation errors remain within the tolerable limits of ADMM-FFT. The distance between the actual result of the FFT operation and a value stored in the memoization database is measured using cosine similarity ($CS$), as defined in Eq. \ref{eq:cs}. 
\begin{equation}
    CS = \frac{K_{query} \cdot K_{db}}{||K_{query}|| \cdot ||K_{db}||}
    \label{eq:cs}
\end{equation}
\noindent where $K_{query}$ and $K_{db}$ are the query key and the key associated with the value, respectively.

We use $\tau$ to control the memoization and reconstruction quality (as mentioned in Section \ref{sec:dm}), specifically, the FFT operation is replaced with the retrieved value when $CS > \tau$.

The determination of $\tau$ to maintain the quality of the final output of ADMM-FFT is empirical and based on domain science. For example, for biological tissues and high-density materials with complex structures, $\tau = 0.95$ is effective to capture fine details (e.g., signal traces between 10 to 100 $\mu m$). In contrast, for large-scale materials such as printed circuit boards (PCBs) and low-density composites, $\tau = 0.9$ is effective in reconstructing large features (0.15-0.3 $mm$). 



\textbf{Convergence.} \textcolor{dong}{Applying memoization may impact the convergence of ADMM-FFT; as a result, it may take a larger number of iterations to reach the required quality, compared to ADMM-FFT without memoization. We observe that (1) the impact of memoization on the convergence is also controlled by $\tau$; (2) the selection of $\tau$ to maintain the quality and convergence as the original ADMM-FFT is often aligned, i.e., $\tau$ that can maintain the quality can also maintain the convergence.}



%% file: text/scale_up_scale_out.tex
\section{Scalable ADMM-FFT}

\begin{figure*}[!t]
\vspace{-13pt}
    \centering      
    \includegraphics[width=2\columnwidth]{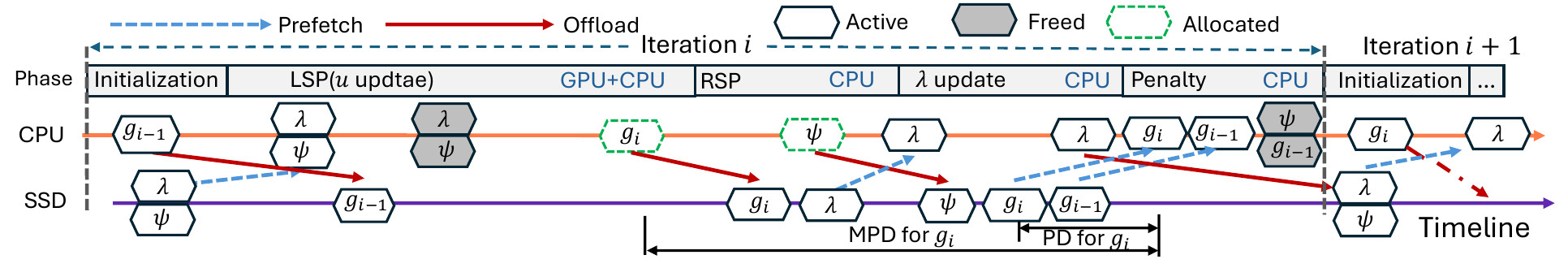}  
        \caption{ADMM-Offloading in the iteration $i$ of ADMM-FFT.}
        \label{fig:offload}
        \vspace{-13pt}
\end{figure*}

We discuss how to  enable ADMM-FFT on larger input problems. 

\subsection{ADMM-Offload}
\label{subsec:ADMM-Offload}

\name offloads memory-consuming variables to SSD to save the CPU memory. An iteration of ADMM-FFT has four execution phases: LSP computation, RSP computation, lambda update, and penalty update. Each phase accesses a set of variables. To maximize CPU memory saving, the variables in a phase are immediately offloaded to SSD once they are not accessed by the same phase. 
Also, to hide data movement overhead, a variable is prefetched from SSD to the CPU memory before the variable is accessed by a phase. We use the following constraints to decide variables for offloading. 

\textbf{Variable selection constraints.} We select a variable as an offloading \textit{candidate} if that variable does not have pointer aliases, and hence offloading and prefetching do not involve extensive changes to ADMM-FFT. Those variables account for more than 80\% of memory consumption. 

Furthermore, we use the following constraints for variable selection. 
We define a term,  \textit{prefetch distance (PD)} of a variable, as follows: the PD of a variable is
the time duration from the moment the prefetch  happens to the moment the execution phase consumes the variable. We define a term, \textit{maximum prefetch distance (MPD)} of a variable, as follows:  the MPD is defined in terms of execution phase, and the MPD of a variable is the time interval between the last access to the variable in the execution phase $k$ to the first access to the same variable in the next execution phase $k+1$. 

Given the above definition, we have the following constraints to select variables for offloading.
\vspace{-0.3em}
\begin{enumerate}
    \item Prefetch must happen \textit{after} offloading; 
    \item If the PD for a variable is 0, then that variable is not offloaded;
    \item Offloading time must be smaller than the MPD;
    \item Prefetch for an execution phase must be finished before that execution phase starts.
\end{enumerate}
\vspace{-0.3em}
The constraint (1) prevents the concurrent occurrence of offloading and prefetching to avoid data race. The constraint (2) disables offloading when the PD is too small. The constraint (3) requires that the time for offloading a variable must be smaller than the time interval between the last access and next access. The constraint (4) is used to prevent invalid accesses.

The above constraints do not guarantee that offloading and prefetch are completely hidden from the critical path. It is possible that prefetch for an execution phase is not finished when that execution phase is about to start. In that case, the execution phase is delayed, hence exposing the prefetch to the critical path. Overall, there is a tradeoff between prefetch overhead and memory saving.

\textbf{Variable selection for offloading.} To determine which variable should be selected and to balance the above tradeoff, we introduce a metric, $MT$. For an offloading/prefetch plan (potentially including multiple variables), $MT$ is a multiplication of memory saving ($M$) and the inverse of performance loss ($1/T$, where $T$ is the performance loss). A larger $MT$  indicates larger memory saving or smaller performance loss (i.e., smaller $T$).  We evaluate various offloading/prefetch plans, following the four constraints, and select the one leading to the largest $MT$. 

\textbf{Forming offloading/prefetch plans.} For each offloading candidate (a variable), we measure its size and identify its first and last accesses in each execution phase. This requires profiling only a single ADMM-FFT iteration and can be automated.

During an execution phase, a variable can be offloaded after its last access; however, if there is insufficient time to offload due to its next access in the subsequent execution phase, the offload operation is skipped.
In each phase, a variable, if offloaded, can be prefetched for a future phase to access. The prefetch happens right after the offloading is done, when the prefetch time is too short to finish; or the prefetch happens when there is just enough time to hide the prefetch overhead before the future phase accesses it. 

An offloading/prefetch plan determines when one or more variables are offloaded or prefetched in each execution phase. Given the offline profiling results, \name can estimate memory saving in each execution phase and performance loss caused by prefetching for each offloading/prefetch plan. Hence, \name can calculate $MT$.

After the above variable selection process, we choose the intermediate variables, $\psi$, $\lambda$ and $g$ for offloading and prefetch. Figure~\ref{fig:offload} depicts the offloading/prefetch plan.

\textcolor{dong}{
\textbf{Why not LRU?} Besides using the above approach to decide variable offloading, we can use a LRU policy: given a limited GPU memory capacity, we offload a variable based on LRU when we need to fetch another variable from SSD to GPU.  However, using LRU has problems: (1) LRU only decides when to offload, but cannot decide when to prefetch; (2) LRU cannot decide whether offloading a variable (and later fetching) causes large performance loss.}

\textcolor{dong}{ADMM-Offload solves the above problem in LRU, and immediately offloads selected variables when they are not used in the current execution phase. Our evaluation shows that ADMM-Offload outperforms LRU-based offloading by 40.5\% on average.}

\subsection{Scaling across GPUs}
\label{subsec:Data Partitioning}
The ADMM-FFT's input can be partitioned into independent chunks using the existing work \cite{nikitin2024laminography,osti_1860447}. Specifically, ADMM-FFT input is a 3D array, in which one of the dimensions is used for partitioning the dataset into chunks. 


\textcolor{dong}{To scale across GPUs, \name evenly distributes the chunks between GPUs within and across nodes. Since the FFT operations work on the chunks generated along different directions---either vertically or horizontally, they can happen without dependency. The original ADMM-FFT cannot run on multiple GPUs/nodes. \name is the first that enables ADMM-FFT to run across multiple GPUs/nodes.}

%% file: text/evaluation.tex
\section{Evaluation}
\label{sec:eval}


\subsection{Evaluation Setup} 
\textbf{Evaluation platform.} We use the Polaris supercomputer at Argonne Leadership Computing Facility. Each node on Polaris is equipped with a single AMD EPYC 7543P processor featuring 32 Zen3 cores (64 hardware threads) operating at 2.8 GHz. Each node has 512 GB DDR4 RAM and four NVIDIA A100 GPUs, where each GPU has 40 GB HBM2 memory and interconnected to others via NVLink. Further, each node also has two local NVMe SSDs providing 3.2 TB capacity. The network across nodes is based on dual HPE Slingshot 11 and its bidirectional injection bandwidth is 200 Gb/s.  
We use CUDA 12.4.1 and its CuFFT library. The chunk size and $\tau$ are 16 and 0.92 by default, respectively. 

\textbf{Evaluation metric.}
To study memoization impact on ADMM-FFT outcome (i.e., reconstruction accuracy), we use this metric.
\begin{equation}
E=\frac{\left\|R_{comp}-R_{\text {LB }}\right\|_\mathcal{F}}{\left\|R_{comp}\right\|_\mathcal{F}}
\end{equation}
$E$ evaluates the relative error between the reconstructed object $R_{comp}$ computed using the original ADMM-FFT and the reconstructed object $R_{LB}$ computed using \name based on memoization, where $\left\|\cdot \right\|_F$ is the Frobenius norm \cite{Zheng2016_quantum_tomo,Han2016}. 

We define the memoization accuracy as follows. 
\begin{equation}
Accuracy = 1 - E
\end{equation}

\noindent A higher accuracy value indicates that the memoization-based computation result is close to that of the original computation, hence  preserving the fidelity of the computed results. 

\textbf{Datasets.} We use 3 datasets, representing small, medium, and large inputs: $1K \times 1K \times 1K$ (or $1K^3$), $1.5K \times 1.5K \times 1.5K$ (or $(1.5K)^3$), and $2K \times 2K \times 2K$ (or $(2K)^3$).  By default, we use the first two.

\begin{figure}[t!]
    \centering      
    \includegraphics[width=1.0\columnwidth]{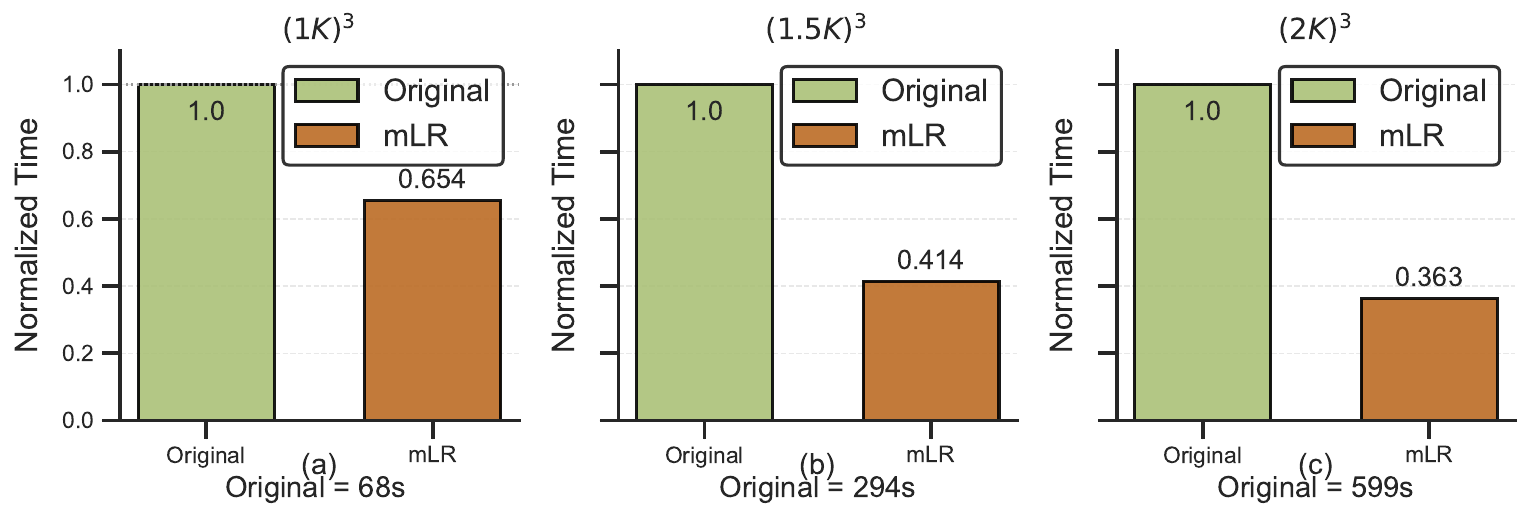}  
        \caption{Overall performance on three datasets.  The performance is normalized to that without using \name.}
        \label{fig:eval_opperation}
           \vspace{-15pt}
\end{figure}

\subsection{Overall Performance}

We use a pair of a compute node (using one GPU and 512 GB memory) and a memory node (using 512 GB memory and up to 1.5 TB SSD) to evaluate the contribution of each optimization method to the overall performance. We employ three different datasets and 60 iterations in ADMM-FFT. We apply all techniques and show the results in Figure~\ref{fig:eval_opperation}. Performance is normalized to that of the original computation without using any performance optimization. 

In general, using \name leads to 34.6\%-65.4\% performance improvement (52.8\% on average) for the three datasets compared to the original ADMM-FFT implementation.


\subsection{Operation Cancellation and Fusion}

\begin{figure}[t!]
    \centering      
    \includegraphics[width=1.0\columnwidth]{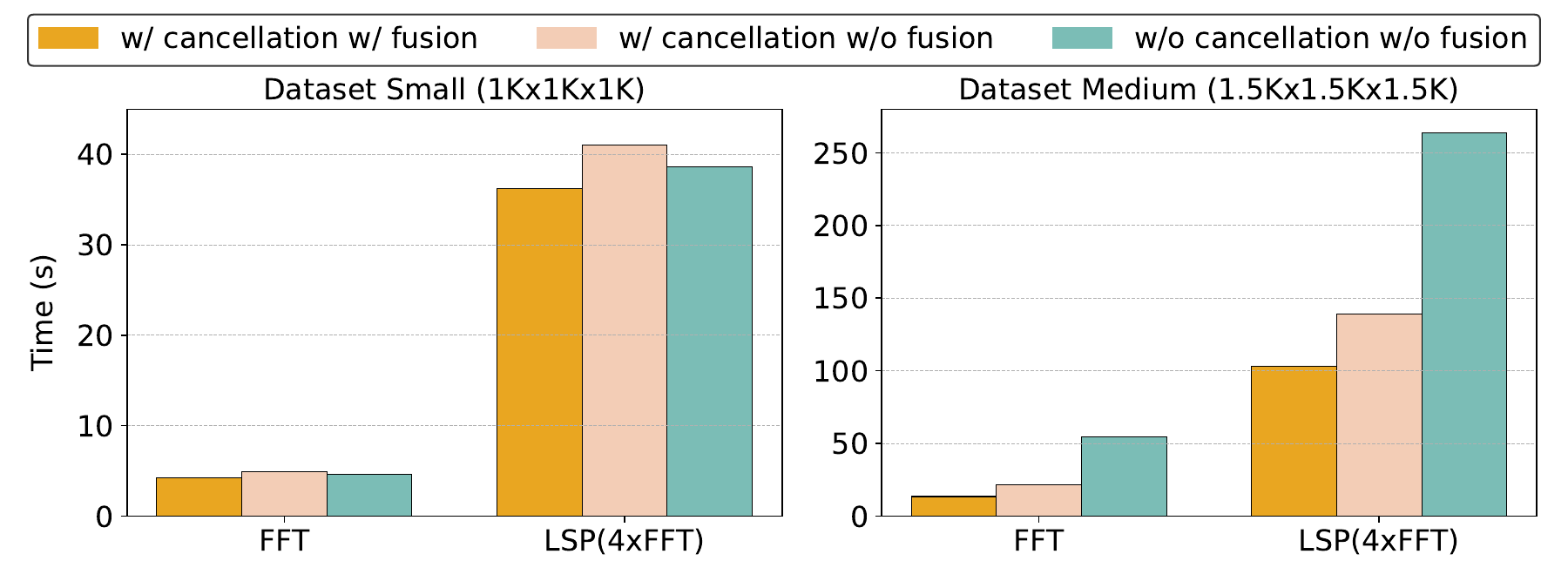}  
        \caption{Comparison of FFT (one forward pass and one adjoint pass) and LSP computation (with $N_{inner} = 4$ ) under different strategies of operation cancellation and fusion.}
        \label{fig:exp_cancel_fusion}
        \vspace{-10pt}
\end{figure}

In this section, we evaluate the effect of operation cancellation and fusion, while disabling other optimizations. 
Figure~\ref{fig:exp_cancel_fusion} presents the results for FFT forward and adjoint operators (Lines  4-8 in Algorithm \ref{alg:lsp_fft_right}), and the whole LSP with $N_{inner} =4$ on a single GPU. 



We have three observations: First, operation cancellation and fusion lead to great performance improvement, compared to the cases without using this technique. For $1K^3$, there are 9.4\% and 7.1\% performance improvements for FFT and LSP respectively; for $(1.5K)^3$, there are 75.3\% and 60.1\% performance improvements for FFT and LSP respectively.  

Second, a larger dataset benefits more from the operation cancellation and fusion (especially cancellation) than a smaller dataset, as shown above. There are two reasons for this: (1) The operation cancellation eliminates redundant data transfers between CPU and GPU, reducing data movement by 1/3, which is especially beneficial for larger datasets; (2) a larger dataset can fully utilize thread-level parallelism for data copy and CPU computation.

Third, the operation cancellation without fusion leads to performance improvement in the medium dataset, but not the small dataset, compared to the cases without using the cancellation and fusion. In particular, for $1K^3$, there is 5.6\% performance loss, but for $(1.5K)^3$, there is 61\% performance improvement.  This is because after the cancellation, the subtraction ($\hat{d}' - \hat{d}$, shown in Line 8 at Algorithm~\ref{alg:lsp_fft_right}) happens in the frequency domain using \texttt{COMPLEX64} on the CPU, which is computationally more expensive than using  \texttt{INT64} operations in the spatial domain. For the smaller dataset, this overhead outweighs the benefits of the cancellation, whereas, for the larger dataset, gain from the cancellation is greater.

\begin{figure}[t!]
    \centering      
    \includegraphics[width=1.0\columnwidth]{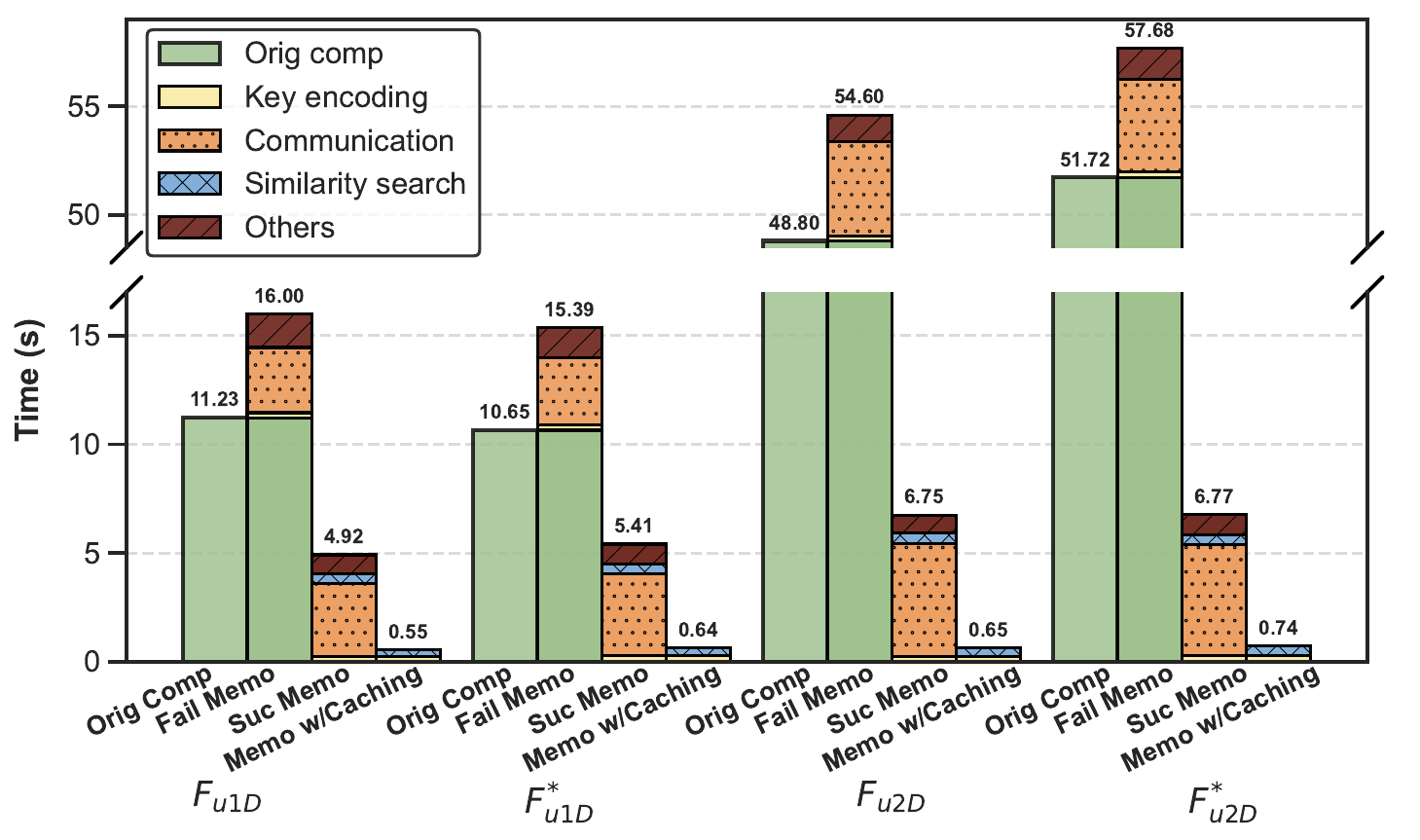}
        \caption{Memoization breakdown for one forward and adjoint on the input,  where Orig comp = orgianl computation, Fail Memo = failed memoization, Suc Memo = successful memoization, Memo w/Caching = memoization with caching.}
        \label{fig:memo_breakdown}
        \vspace{-10pt}
\end{figure}

\subsection{Memoization Breakdown Analysis}
We enable the operation cancellation and fusion optimizations, use one node with one GPU, and reconstruct a $1K^3$ volume. Figure~\ref{fig:memo_breakdown} shows the memoization performance for processing one chunk within one iteration of the main loop (as shown in Algorithm~\ref{alg:lsp_fft_right}).  When applying the memoization, there are three possible cases: (1) \name cannot find any matched value in the memoization database, and has to perform the original FFT computation, followed by a remote insertion process. In Figure~\ref{fig:memo_breakdown}, the second bar in each group represents this case. 
(2) \name finds a matched value in the memoization database without using the local cache, which is the 3rd bar within the groups; and (3) \name finds a matched value in the local cache, which is the 4th bar within the groups. We have 3  observations. 

First, the case (1) has small performance difference (less than 2.5\%) compared to the original computation---i.e., it does not bring performance benefit but the performance loss is small. Having the small performance loss is because the most part of the insertion process, including communication and remote database access, are overlapped with the next iteration's computation. Although the key encoding is exposed to the critical path, its overhead is trivial (less than 1\%). 

Second, the case (2) brings performance benefit, even though \name has to interact with the remote memory node without using the local cache. The performance benefit is 55\% for $F_{u1D}$ and $F^{*}_{u1D}$, and 88\% for $F_{u2D}$ and $F^{*}_{u2D}$, relative to the original computation. By replacing the FFT computation, the memoization significantly reduces the computation time, which also compensates the communication and key encoding overheads. 

Third, the case (3) brings performance benefit, compared to (2). For example, for $F_{u1D}$, the local cache brings 85\% performance improvement. The cache effectively reduces remote database operations (search, similarity comparison, and communication).

\textbf{Distribution of the three cases.} 
With our default setting ($\tau = 0.92$) and the dataset $(1.5K)^3$, the cases 1, 2, and 3 account for 53\%, 19\%, and 28\% of all iterations, respectively. Memoization is capable of reducing computation for chunks involving $F_{u*D}$ by an average of 47\%. In the other two datasets, we see the similar distribution. 



\begin{figure}[t!]
    \centering      
    \includegraphics[width=0.95\columnwidth]{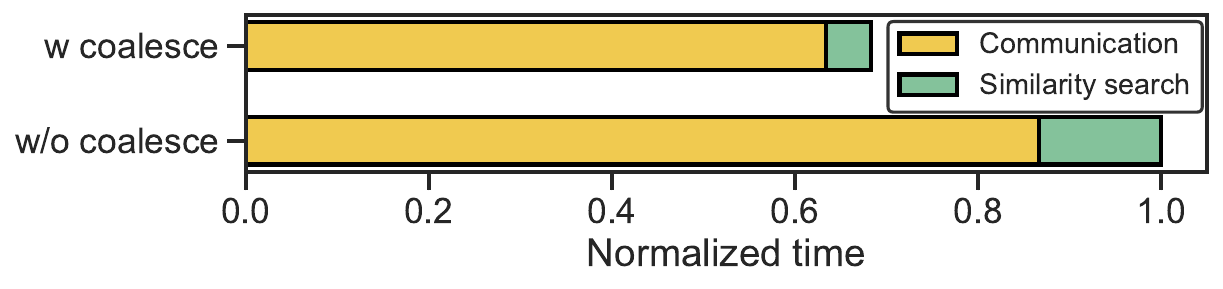}
        \caption{Average performance w and w/o key coalesce for one chunk in an iteration of ADMM-FFT, 
 normalized by that w/o key coalesce.}
        \label{fig:key_coalesce}
        \vspace{-15pt}
\end{figure}

\textbf{Key coalesce} is used to optimize the payload size (Section~\ref{sec:op_payload_size}). Figure \ref{fig:key_coalesce} shows the performance with and without key coalesce, using the dataset $1K^3$. The figure reports the communication time exposed to the critical path and similarity search time in the memoization database. Key coalescing improves performance by 25\%, due to better bandwidth usage and faster batched search for similarity.

\subsection{FFT Cache}
\label{sec:fft_cache}

\begin{figure}[t!]
    \centering      
    \includegraphics[width=1.0\columnwidth]{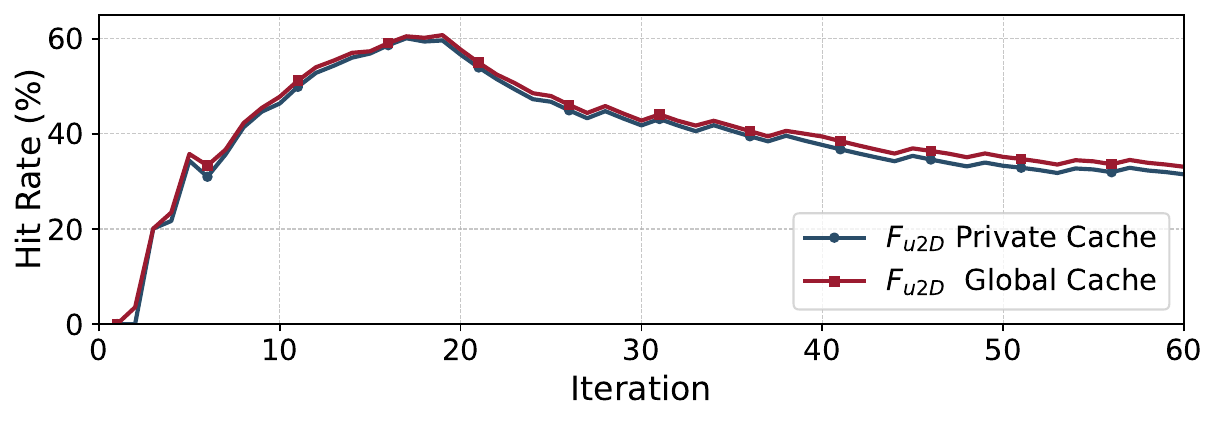}  
        \caption{Hit rate of the private and global caches for $F_{u2D}$.}
        \label{fig:usfft2d_cache}
        \vspace{-15pt}
\end{figure}

We study the hit rate of different caching policies for memoization. We compare the hit rate of the private cache employed by \name and a global cache. The global cache allows cross-location data sharing, which means chunks can be shard across different locations. Figure~\ref{fig:usfft2d_cache} shows the results for one operation, $F_{u2D}$. 

Figure~\ref{fig:usfft2d_cache} shows that the private and global caches show similar hit rates throughout the iterations. Using the global cache does not provide performance benefits, whereas the private cache brings 85.7\% performance gain. The main reason for this improvement is that the private cache only needs to perform similarity comparison for once (there is only one item in the private cache), while the global cache has to perform 64 for the $1K^3$ dataset.

\begin{figure}[t!]
    \centering      
    \includegraphics[width=1.0\columnwidth]{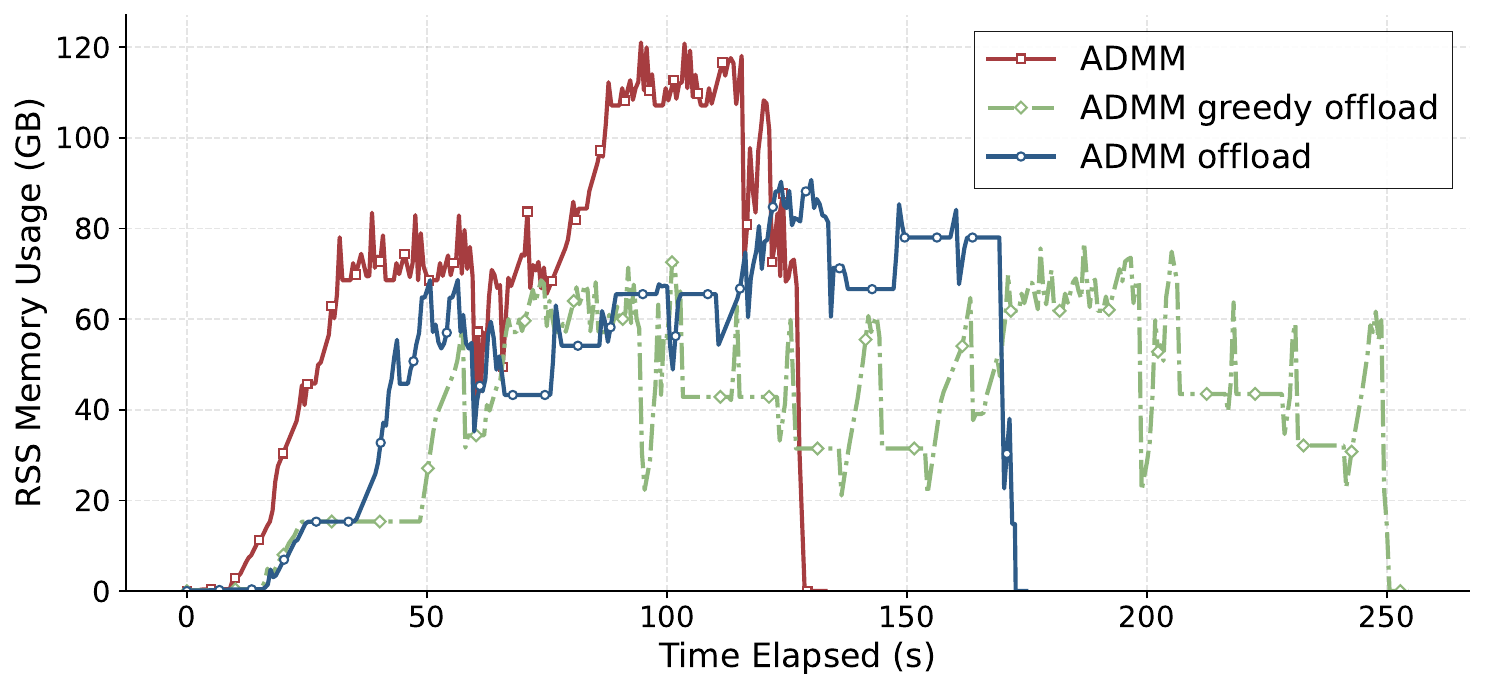}  
        \caption{ADMM-offload with execution time and GPU memory consumption.}
        \label{fig:offload_aval}
        \vspace{-5pt}
\end{figure}



\subsection{ADMM-Offload}
We study ADMM-Offload on a single node with a single GPU and the dataset $1K^3$. To focus on ADMM-Offload, we do not apply memoization. We evaluate three cases: (1) ADMM without offload, (2) ADMM with greedy offload, and (3) ADMM-Offload. The greedy offload is a strategy that immediately offloads the four variables that use the most memory upon generation and fetches them on demand when needed. As a result, the overhead of data offload is largely exposed to the critical path. Figure \ref{fig:offload_aval} shows the memory consumption at runtime and execution time for the three cases. 

Figure \ref{fig:offload_aval} shows that without any offloading, the peak memory consumption is over 121 GB. In contrast, ADMM with greedy offload saves memory by 42\% but loses performance by 81.5\% ($MT=0.51$). ADMM-Offload reduces the maximum memory consumption to 86 GB, which saves memory by 29\% but loses performance by only 21\% ($MT=1.38$). ADMM-Offload strikes a better balance between memory saving and avoiding performance loss.



\subsection{Scalability}

\begin{figure}[t!]
    \centering      
    \vspace{-10pt}
    \includegraphics[width=1.0\columnwidth]{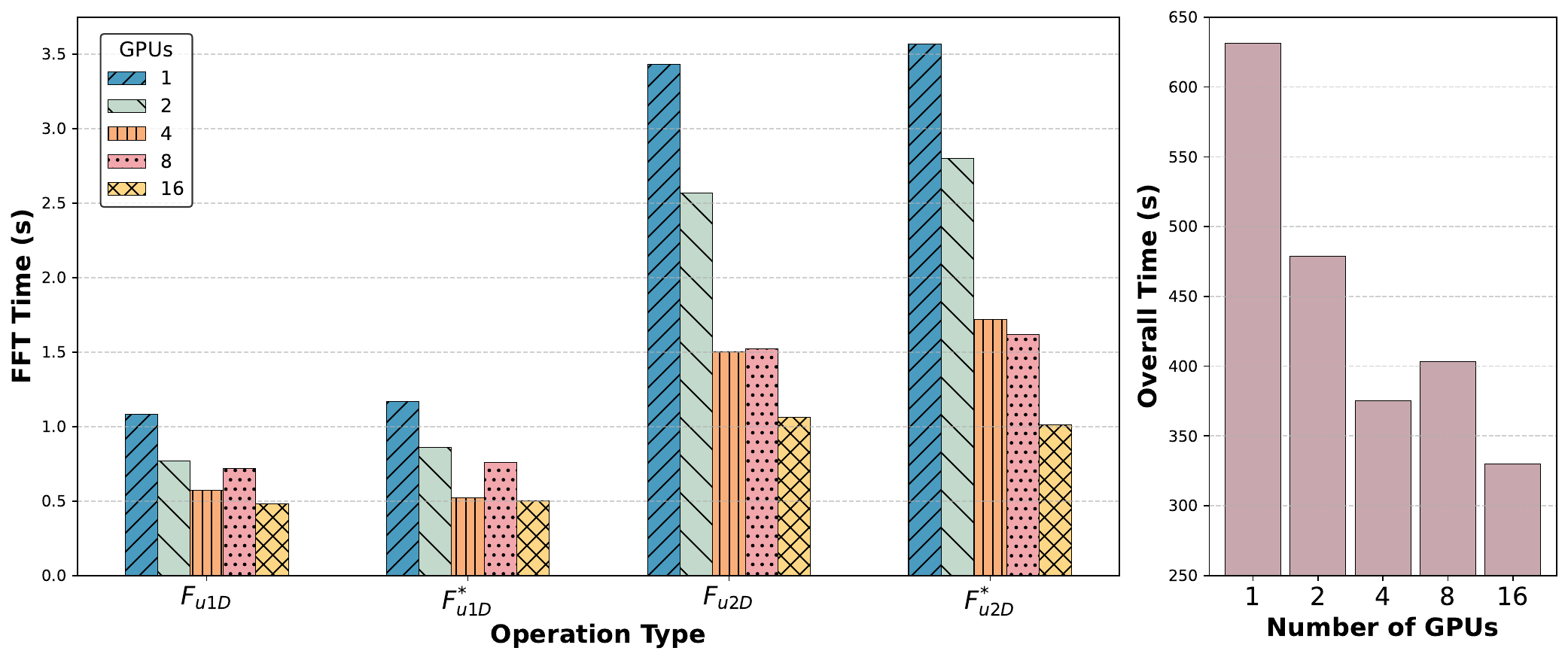}  
        \caption{Scalability of the FFT operations and ADMM-FFT over multiple GPUs on the dataset $1K^3$.}
        \label{fig:small_scale}
        \vspace{-11pt}
\end{figure}



We evaluate the scalability of FFT operations using datasets $1K^3$. \textbf{Performance.} Figure \ref{fig:small_scale} shows the performance of the FFT operations and overall execution time across different number of GPUs, where inter-node reconstruction is observed after 4 GPUs (each node has 4 GPUs). We have two observations.

\begin{figure}[t!]
    \centering      
    \vspace{-13pt}
    \includegraphics[width=1.0\columnwidth]{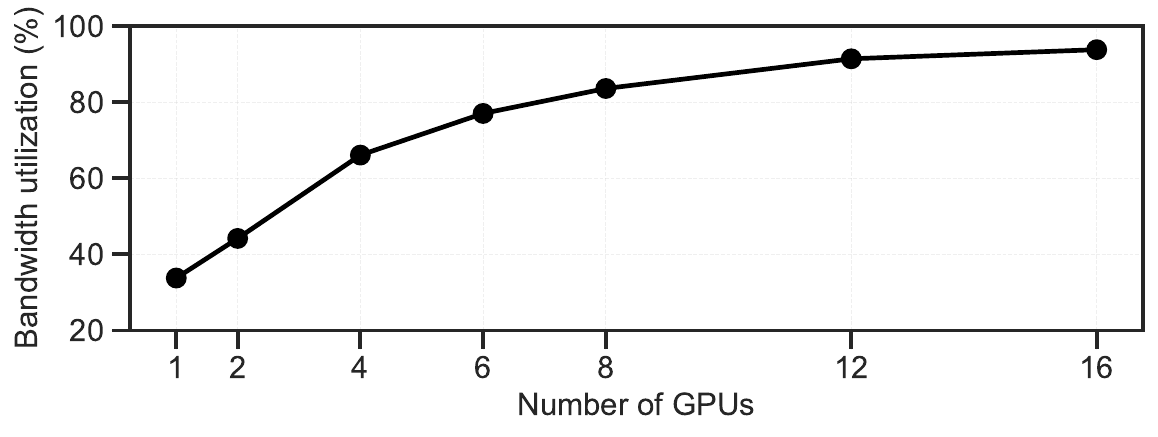}     
    \vspace{-17pt}
        \caption{The interconnect bandwidth utilization with different number of GPUs (each node has four GPUs).}
        \label{fig:bandwidth_nodes}
    \vspace{-13pt}
\end{figure}

\textit{First}, using more GPUs leads to better performance. For example, on the dataset $1K^3$, the execution time of the operation $F_{u1D}$ decreases from \textcolor{check}{1.1s (1 GPU) to 0.5s (16 GPUs) with $2.2\times$ speedup}. The speedup is not linear due to the inter-GPU communication, data partition, and processing. \textit{Second}, scaling from 4 GPUs to 8 GPUs, there is a noticeable diminishing return in performance. For example, increasing from 2 GPUs to 4 GPUs brings 1.36x speedup in overall performance, but increasing from 4 GPUs to 8 GPUs brings minor performance loss (1\%). Such diminishing performance return comes from the inter-node communication: the configurations with more than 4 GPUs involve multiple compute nodes.

\begin{figure}[t!]
    \centering     \includegraphics[width=1.0\columnwidth]{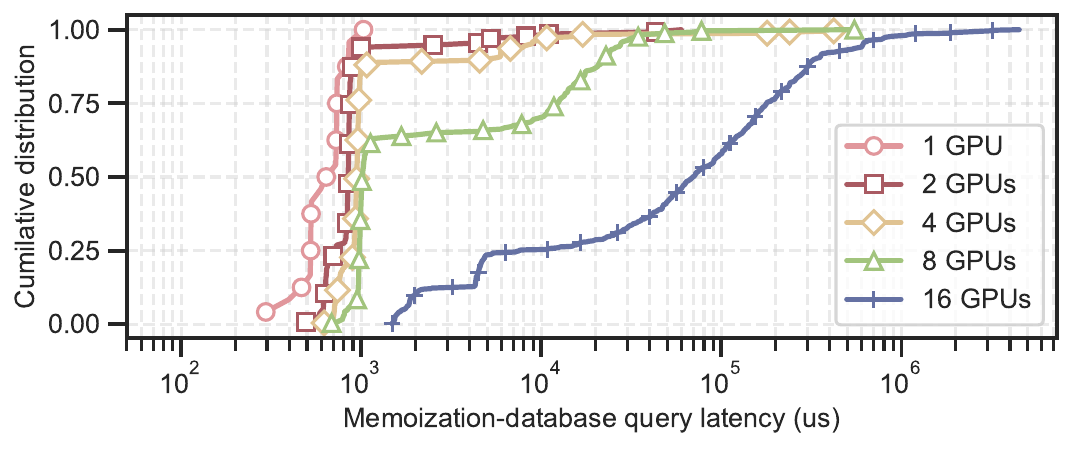}  
       \vspace{-14pt}
        \caption{The latency distribution under contention with different number of compute nodes.}
        \label{fig:latency_nodes}
    \vspace{-15pt}
\end{figure}


\textbf{Interconnect bandwidth analysis.} \textcolor{dong}{Increasing the number of GPUs (and nodes accordingly) can create a performance bottleneck in the interconnect between the compute nodes and memory node.  Figure \ref{fig:bandwidth_nodes} depicts our study using one memory node. When the number of GPU is 12 (3 nodes) or more, the interconnect bandwidth is close to the peak bandwidth, creating a performance bottleneck.}


\textbf{Latency distribution.} \textcolor{dong}{We further study the impact of increasing GPUs on the memoization-database query latency. We use one memory node. Figure~\ref{fig:latency_nodes} shows the cumulative distribution of query latency. In general, as the number of GPUs increases, the latency distribution shifts right. As the number of GPUs is 16, the latency distribution spreads further, and 43\% of the queries have latency longer than 100,000 $\mu s$, indicating contention on the interconnect bandwidth and/or memory bandwidth.}

\subsection{Reconstruction Accuracy and Convergence}
We study the impact of memoizatoin on the reconstruction accuracy by changing $\tau$. We use the same number of iterations (60) for all cases. Table~\ref{tab:Accuracy} shows the results using the dataset $1K^3$. We have two observations. \textit{First}, using a larger $\tau$ leads to higher accuracy. This is expected as a larger $\tau$ has a more strict requirement on the similarity search. \textit{Second}, when $\tau$ is larger than 0.92, the accuracy is at least 0.94. Considering the target reconstruction quality and features, this accuracy threshold was found satisfactory by the domain scientist in our evaluation; thus, we set $\tau = 0.92$ in our evaluation. Note that this threshold can be adjusted depending on the target phenomena or sample morphology, making our method applicable to a wide range of use cases.


\begin{table}[!t]
   \caption{Impact of memoization on the reconstruction accuracy. We use the dataset $1K^3$.} 
   \vspace{-10pt}
   \label{tab:Accuracy}
   \small
   \centering
   \begin{tabular}{l|c|c|c|c|c|r}
   \toprule
   \textbf{Threshold $\tau$} & \textbf{0.86} & \textbf{0.88} & \textbf{0.90} & \textbf{0.92} & \textbf{0.94} & \textbf{0.96} \\ 
   \midrule
  $Accuracy$ & 0.691 & 0.808  & 0.901 & 0.946  & 0.958  & 0.973\\ 
   \bottomrule
   \end{tabular}
    \vspace{-10pt}
\end{table}

\begin{figure}[t!]
    \centering 
  \includegraphics[width=1.0\columnwidth]{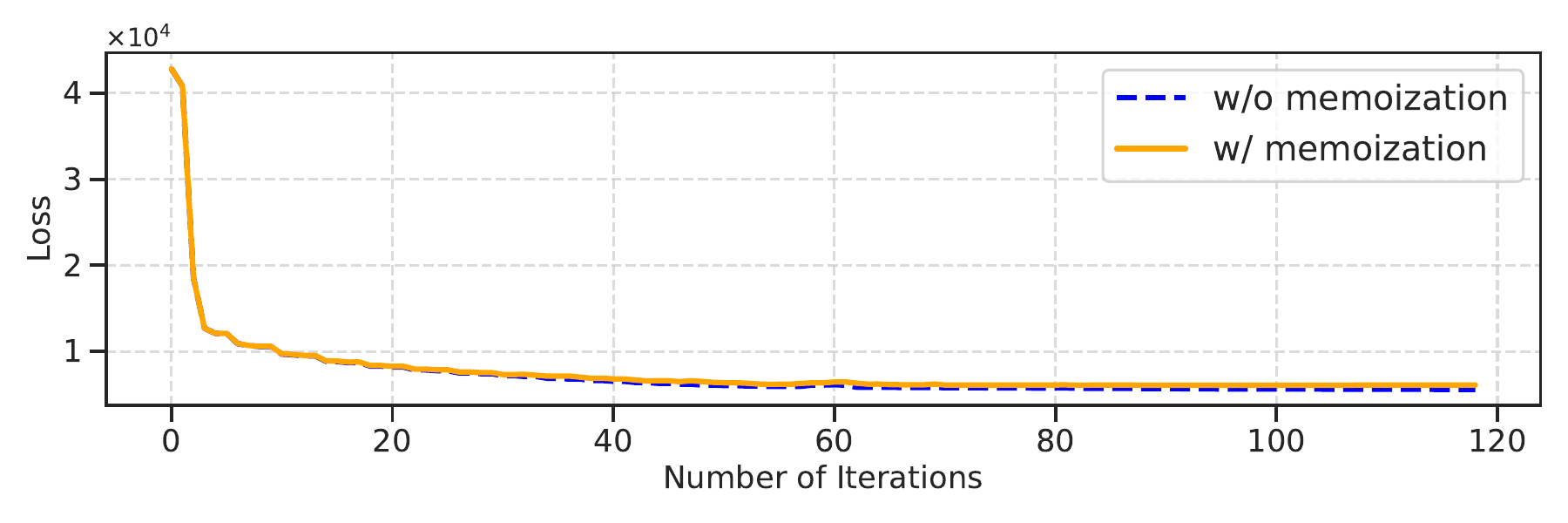}
        \vspace{-10pt}
        \caption{Convergence with and without memoization.}
        \label{fig:convergence_loss}
        \vspace{-15pt}
\end{figure}


\textcolor{dong}{We further study the convergence loss curve using the same dataset and $\tau$. Figure~\ref{fig:convergence_loss} shows the results. We observe that the convergence loss curves with and without memoization are pretty close. With memoization, ADMM-FFT does not need extra iterations to reach the similar convergence as without memoization. We see the same trend in other datasets (i.e., $(1.5K)^3$ and $(2K)^3$).}



%% file: text/related_work.tex
\section{Related Work}
\textbf{Computed tomography (CT)} is a type of non-destructive imaging technique that allows the investigation of an object’s interior structure \cite{kalender2006x}.  It reconstructs a 3D image from a series of x-ray projections taken from different angles around the object. As a well-established technique, CT has been widely used in various fields \cite{wevers2018applications, du2016comparison, wang2020non, Schabowicz2019NonDestructiveTO, gupta2022advances, chen2021novel, sluimer2006computer, salvo20103d, GPUiterXiaodong2019,banerjee2025inpainting}.
However, traditional CT  may fail to capture sufficient 3D information  for samples that are extremely thick or thin along one dimension \cite{zuber2017augmented}.

\textbf{Laminography reconstruction (LR)} is another non-destructive imaging technique that addresses some of the limitations of traditional CT, such as limited X-ray penetration and geometrical constraints. As a result, LR has been increasingly applied in scenarios requiring high-resolution imaging of planar objects, including electronic circuit boards, composite materials and biological tissues\cite{zhou1996computed, gondrom1999x, o2016recent, verboven2015synchrotron, helfen2013nano}. Various optimizations for laminography have been explored \cite{nikitin2024laminography, o2016comparing, ji2024fusional, voropaev2016direct, myagotin2013efficient, venkatakrishnan2017model,lamino_HPCI24}. For example, 
Nikitin at al. \cite{nikitin2024laminography}  accelerate LR on multi-GPUs to efficiently handle large datasets. Voropaev et al. \cite{voropaev2016direct} derive a Fourier-based reconstruction equation designed for circular laminography (CL) scanning geometry,  reducing reconstruction time while preserving image quality.
Our study differs from existing works by focusing on optimizing computational efficiency, memory utilization, and scalability in LR. 

\textbf{Memoization} optimizes computation efficiency by saving and reusing previous computational results. It has been applied across various fields \cite{silfa2019neuron, liu2019axmemo, xie2021md, feng2023attmemo, essofi2023memoization, mangrulkar2024optimization, mangrulkar2023optimizing, steiner2023model, liu2019axmemo, keramidas2015clumsy, rahimi2013spatial, xie2021md,riann2020}.  
In AI/ML optimization, 
Silfa et al. \cite{silfa2019neuron} apply fuzzy memoization to recurrent neural network (RNN), reducing training computation by more than 24.2\%.
Yuan et al. \cite{feng2023attmemo} employ  memoization to substitute expensive self-attention computations for accelerating transformers. Steiner et al. \cite{steiner2023model} represent neural networks as directed acyclic graphs (DAGs) and leverage memoization to cache vertex reachability query results,  which is critical for efficient dependency resolution and topological optimization.
For the architecture and compiler designs, Liu et al. \cite{liu2019axmemo} employ a two-level memoization lookup to efficiently alleviate execution overhead of code segments. 

\textbf{Tiered memory.} Tiered memory \cite{ dynn-offload, ppopp23:merchandiser, atc21:zerooffload, ics21:warpx, ics21:athena, ppopp21:sparta, hpca21:ren, neurips20:hm-ann, Wu:2018:RDM:3291656.3291698, unimem:sc17, cluster17:huang, shuo:cluster17,atc24_flexmem,RecMG_hpca25} orchestrate heterogeneous memory components with varying characteristics in terms of latency and cost \cite{buffulo_hpca25,RecMG_hpca25,ScientificNVM, HPCnet23,ribbon, InfiniStore,infinicache,lan2025zenflow}. At its core, \name represents a memory tiering approach that pioneering applies this paradigm specifically to HPC environments and X-ray image reconstruction workloads.

%% file: text/conclusion.tex
\section{Conclusions}
We introduce \name that uses memoization to accelerate ADMM-FFT and scale it across GPUs within and across nodes. The design of \name is based on our observations on workload characterization (e.g., the appearance of similar operations, and variable liveness across execution phases). Using \name, we enable larger input problems on ADMM-FFT with limited memory and bring 52.8\% performance improvement on average (compared to using the traditional methods), which brings new opportunities for scientific discovery. 


%% file: text/acknowledgment.tex
\begin{acks}

\textcolor{check}{This research used resources of the Advanced Photon Source (APS) and Argonne Leadership Computing Facility (ALCF), U.S. Department of Energy (DOE) Office of Science user facilities, and is based on work supported by Laboratory Directed Research and Development (Project Number: 2023-0104) funding from Argonne National Laboratory, provided by the Director, Office of Science, of the U.S. DOE under Contract No. DE-AC02-06CH11357.} \textcolor{check}{This work was also partially supported by U.S. National Science Foundation (2104116, 2316202 and 2348350). We would like to thank the anonymous reviewers for their feedback on the paper.}
\end{acks}